\documentclass[11pt]{article}
\usepackage{amsmath, amsfonts, amssymb}
\usepackage[margin=1in]{geometry}
\usepackage{csvsimple}
\usepackage{graphicx}
\usepackage{natbib}
\usepackage[figuresright]{rotating}
\usepackage{setspace}
\usepackage{caption}
\usepackage{subcaption}
\newcommand{\mc}{\multicolumn}

\begin{document}
\bibliographystyle{ECA_jasa}

\title{Microsimulation Model Calibration using Incremental Mixture Approximate Bayesian Computation}

\author{Carolyn Rutter$^{1}$, Jonathan Ozik$^{2}$, Maria DeYoreo$^{1}$, and Nicholson Collier$^{2}$ \\
	   $^{1}$RAND Corporation\\
	   $^{2}$University of Chicago and Argonne National Laboratory\\
	   %\texttt{crutter@rand.org, jozik@anl.gov, mdeyoreo@rand.org, ncollier@anl.gov}
	   %J. Ozik$^{*}$\email{email1aa@address.com}\\
	  % Building, Institute, Street Address, City,
		   }
\maketitle

\begin{abstract}
Microsimulation models (MSMs) are used to inform policy by predicting population-level outcomes under different scenarios. MSMs simulate individual-level event histories that  mark the disease process (such as the development of cancer) and the effect of policy actions (such as screening) on these events. MSMs often have many unknown parameters; calibration is the process of searching the parameter space to  select parameters that result in accurate MSM prediction of a wide range of targets. We develop Incremental Mixture Approximate Bayesian Computation (IMABC) for MSM calibration, which results in a simulated sample from the posterior distribution of model parameters given calibration targets. IMABC begins with a rejection-based ABC step, drawing a sample of points from the prior distribution of model parameters and accepting points that result in simulated targets that are near observed targets. Next, the sample is iteratively updated by drawing additional points from a mixture of multivariate normal distributions and accepting points that result in accurate predictions. Posterior estimates are obtained by weighting the final set of accepted points to account for the adaptive sampling scheme.  We demonstrate IMABC by calibrating CRC-SPIN~2.0, an updated version of a MSM for colorectal cancer (CRC) that has been used to inform national CRC screening guidelines.
\end{abstract}

%
%  Please place your key words in alphabetical order, separated
%  by semicolons, with the first letter of the first word capitalized,
%  and a period at the end of the list.
%

%\keywords{Adaptive ABC; Agent-based models; Colorectal cancer; High performance computing}

\section{Introduction}
Microsimulation models (MSMs) are used to inform policy by predicting population-level outcomes under different policy scenarios. MSMs are characterized by simulation of \textit{agents} that represent individual members of an idealized population of interest. For each agent, the model simulates event histories that catalog landmarks in the disease process. 
%MSMs use mathematical models to approximate disease processes at the individual-level, in contrast to multi-scale models that describe processes at the individual and cellular levels \citep{deisboeck2011}. 
In general, %the individual-level 
disease processes modeled are not directly observable, though outcomes from these processes may be observed. 
For example, the process of developing colorectal cancer (CRC) cannot be observed, but the prevalence of both precursor lesions (adenomas) and preclinical (asymptomatic) CRC can be estimated from screening trials, and CRC incidence can be observed from national registry data. 

Model calibration involves selecting parameter values that result in model predictions that are consistent with observed data and expected findings. 
%Through calibration, MSMs synthesize information from targets derived from randomized controlled trials, observational studies, and expert opinion. 
Once parameters are selected, MSMs can be used to make predictions about population trends in disease outcomes, effectiveness of interventions, and the comparative effectiveness of interventions, especially those without direct empirical comparisons. For example, models have been used to inform U.S. Preventive Services Task Force screening guidelines for breast \citep{mandelblatt2016collaborative}, cervical \citep{uspstfcervical2018}, 
colorectal \citep{knudsen2016estimation}, and lung cancer \citep{de2014benefits} by comparing the effectiveness of different screening regimens. 

MSM calibration involves searching a high dimensional parameter space to predict many targets. Several approaches have been proposed. 
The simplest calibration method involves perturbing parameters one at a time 
and evaluating the goodness of fit to calibration data, but this is only feasible when calibrating a few parameters. Directed searches, such as the Nelder-Mead algorithm \citep{Nelder}, provide a derivative free hill-climb to identify a single best value for each parameter. \citet{Kong} used search algorithms from engineering (simulated annealing and a genetic algorithm) for model calibration. 
%This approach requires defining a goodness of fit score determined by a weighted sum of calibration-specific terms, with user defined weights. 
Bayesian calibration methods estimate the joint posterior distribution of MSM parameters, which
provides information about parameter uncertainty and enables estimation of functions of parameters.
% and can be used for value of information analysis. 
\citet{Rutter2009} used Markov Chain Monte Carlo (MCMC) to simulate draws from the posterior distribution of MSM parameters given calibration targets. However, MCMC can be difficult and costly to apply to MSM calibration and because MCMC is based on a process of sequentially updating draws, it is not easy to parallelize the process to take advantage of modern computing resources.

Approximate Bayesian Computation (ABC) offers an alternative approach to MSM calibration.
ABC is a likelihood-free technique for simulating draws from the posterior distribution 
that approximates likelihood-based algorithms by choosing parameters that produce a close match to data  
rather than calculating the likelihood \citep{Marin2012, Conlan}.
The validity of ABC algorithms, in the sense that they result in samples from the approximate posterior distribution, relies on the validity of the corresponding exact algorithms \citep{Sisson}. 
The idea underlying ABC is simple. For a parameter $\theta$ with prior distribution $\pi(\theta)$ and observed data $y$, we can write the posterior probability as $p(\theta|y) = p(y|\theta)\pi(\theta)$ implying that we can approximate $p(\theta|y)$ by sampling $\theta$ from $\pi(\cdot)$ and retaining only points with $p(y|\theta)\approx 1$.
However, ABC is inefficient and can fail when the parameter space is high dimensional, when there are many calibration targets, or when the prior distributions are very different from the posterior distributions. \citet{McKinley} found that popular ABC variants that improve the algorithm's efficiency were not computationally feasible for calibrating stochastic epidemiological models. We propose an Incremental Mixture ABC (IMABC) approach for MSM model calibration that begins with a basic rejection-sampling ABC step \citep[e.g.,][]{Pritchard} and then incrementally adds points to regions where targets are well predicted. 

In the next sections we describe the CRC-SPIN MSM for the natural history of colorectal cancer (CRC)  (\S \ref{sec:CRCSPIN}), calibration targets used to inform CRC-SPIN model parameters (\S \ref{sec:calibrationdata}), the IMABC calibration approach (\S \ref{sec:iabc}), and results of CRC-SPIN model calibration based on IMABC (\S \ref{sec:results}). We conclude with general remarks about the proposed approach and discussion of future work (\S \ref{sec:discussion}).

\section{Microsimulation Model for the Natural History of Colorectal Cancer}\label{sec:CRCSPIN}
The ColoRectal Cancer Simulated Population Incidence and Natural history model (CRC-SPIN) \citep{Rutter2009,Rutter2010} describes the natural history of CRC based on the adenoma-carcinoma sequence \citep{Muto,Leslie}.  Four model components describe the natural history of CRC: 1) adenoma risk; 2) adenoma growth; 3) transition from adenoma to preclinical cancer; and 4) transition from preclinical to clinical cancer (sojourn time).

CRC-SPIN has been used to provide guidance to the Centers for Medicare and Medicaid Services (CMS) \citep{CMS_CTC} and to inform  U.S. Preventive Services Task Force CRC screening guidelines~\citep{knudsen2016estimation}.
Model validation, based on comparison of model predictions to observed outcomes, 
revealed that while CRC-SPIN predicted many aspects of CRC well 
(including clinically detected cancer, cancer mortality, and the effectiveness of screening), 
it predicted detection of too few preclinical cancers at screening, indicating that the simulated times spent in the preclinical cancer phase (sojourn times) were too short \citep{RutterValidation}. 
In this paper we present CRC-SPIN~2.0, an update to the original CRC-SPIN~1.0. % that updates the sojourn time model and makes other changes to address issues that have arisen during the 10 years that this model has been in use. 
CRC-SPIN~2.0 contains 21 calibrated parameters (Table~\ref{modelsummary}).  Because this is a model recalibration, prior distributions are based on results from the previous calibration of CRC-SPIN~1.0 \citep{Rutter2009}. In this section, we provide an overview of the model. Additional details are provided in Appendix \S \ref{sec:modelappendix} and  online at \texttt{cisnet.cancer.gov} \citep{CISNETwebsite}.

\begin{table}
	\caption{\label{modelsummary}Summary of CRC Microsimulation Model Components. Calibrated parameters associated with the 4 components of the natural history model, including parameter notation, associated equations, prior distributions and posterior estimates (mean and 95\% credible interval). TN$_{[a,b]}$($\mu$,$\sigma$) denotes a truncated normal distribution with mean $\mu$ and standard deviation $\sigma$, restricted to the interval $(a,b)$. U($a$,$b$) denotes a Uniform distribution over $(a,b)$. Refer to section \ref{sec:CRCSPIN} for details of the 4 model components.}
	\begin{tabular}{|l|l|rc|}
		\hline
		&\mc{1}{|c|}{Prior}&\mc{2}{|c|}{Posterior Estimates} \\  
		Component                                    &\mc{1}{|c|}{Distribution} &\mc{1}{c}{Mean} 
		&\mc{1}{c|}{95\% CI}  \\ \hline 
		Adenoma Risk (eqn \ref{adenomariskeqn}) &&&\\
		\hspace*{.1in} Baseline log-risk                    & $A \sim$           TN$_{[-6.7,-6.1]}$(-6.4,0.25) &-6.36 &(-6.65,-6.03) \\
		\hspace*{.1in} Standard deviation, baseline log-risk& $\sigma_\alpha \sim$ U(0.75,1.75)     		   &1.28 &(0.86,1.69)  \\ 
		\hspace*{0.1in} Female                       & $\alpha_1 \sim$    TN$_{[-0.7,-0.3]}$(-0.5,0.1)  &-0.61 &(-0.69,-0.49) \\ 
		\hspace*{.1in} Age effect, age $\in [20,50)$        & $\alpha_{20} \sim$ TN$_{[0.02,0.05]}$(0.04,0.06) &0.041 &(0.033,0.049) \\ 
		\hspace*{.1in} Age effect, age $\in [50,60)$        & $\alpha_{50} \sim$ TN$_{[0.01,0.05]}$(0.03,0.01) &0.028 &(0.012,0.046) \\ 
		\hspace*{.1in} Age effect, age $\in [60,70)$        & $\alpha_{60} \sim$ TN$_{[-0.01,0.05]}$(0.03,0.01)&0.013 &(-0.007,0.039) \\ 
		\hspace*{.1in} Age effect, age $\ge 70$             & $\alpha_{70} \sim$ U(-0.02,0.03)                 &0.008 &(-0.016,0.028) \\ 
		\hline\hline		

		Time to 10mm (eqn \ref{timeto10eqn}) &&&\\
		\hspace*{.1in} Shape, colon                         &$\beta_{1 C} \sim$ U(1.1,5) & 1.32	&(1.12,1.57) \\
		\hspace*{.1in} Shape, rectum                        &$\beta_{1 R} \sim$ U(1.1,5) & 3.30 &(1.68,4.84) \\ 
		\hspace*{.1in} Scale, colon$^*$                     &$\beta_{2 C} \sim$ U(10.7,40)  &38.1&(35.6,39.9) \\
		\hspace*{.1in} Scale, rectum$^*$                    &$\beta_{2 R} \sim$ U(10.7,40)  &16.4 &(13.5,19.3)  \\

		\hline\hline
		\hspace*{.1in} Intercept                            &$\gamma_{0} \sim$ TN$_{[2.6,3.6]}$(3.1,0.25)   &3.23	&(3.07,3.42) \\ 
		\hspace*{.1in} Female (versus male)                 &$\gamma_{1} \sim$ TN$_{[-0.3,0.3]}$(-0.06,0.2) &-0.17 &(-0.26,-0.09)  \\ 
		\hspace*{.1in} Rectal (versus colon)                &$\gamma_{2} \sim$ U(-0.25,0.25)                &-0.07 &(-0.24,0.15)  \\ 
		\hspace*{.1in} Female \& rectal                     &$\gamma_{3} \sim$ U(-0.25,0.25)                &0.12	&(-0.03,0.23)  \\ 
		\hspace*{.1in} Age at initiation                    &$\gamma_{4} \sim$ TN$_{[-0.024,0.002]}$(-0.008,0.004)  &-0.009 &(-0.014,-0.004)  \\ 
		\hspace*{.1in} Female \& age at initiation          &$\gamma_{5} \sim$ U(-0.004,0.004)              &0.001  &(-0.003,0.004)  \\ 
		\hspace*{.1in} Rectal \& age at initiation          &$\gamma_{6} \sim$ U(-0.004,0.004)              &0.000 &(-0.004,0.003)  \\ 
		\hspace*{.1in} Female, rectal, \& age at initiation &$\gamma_{7} \sim$ U(-0.004,0.004)		        &0.000 &(-0.004,0.004)  \\ 
		\hline\hline

		Mean Sojourn Time (eqn \ref{mst})                &&&\\
		\hspace*{.1in} Colon							   &$\tau_{C} \sim$ U(1.5,5.0)   &1.91 &(1.52,2.65) \\ 
		\hspace*{.1in} Rectum				   			   &$\tau_{R} \sim$ U(1.5,5.0)   &2.32 &(1.55,3.55) \\ 
		\hline
	\end{tabular}
	$^*$Scale parameters, $\beta_{2}$, were also restricted to range from $10(-\ln(0.25))^{1/\beta_{1}}$ to $(-\ln(0.0001))^{1/\beta_{1}}$, corresponding to the probability of an adenoma reaching 10mm within 10 years ranging from 0.0001 to 0.25. 
\end{table}

\subsection{Adenoma Risk Model}
The occurrence of adenomas is modeled using a non-homogeneous Poisson process with a piecewise age-effect. We assume zero risk before age 20. 
We focus on CRC in adults because CRC is very rare before age 20, with incidence of about one in 10 million \citep{koh2015care}. %While there is evidence that CRC risk is increasing in younger age groups, the absolute risk remains low (below 1 in 100,000 for 20 to 29 year olds) \citep{siegel2017colorectal}
The $i$th agent's baseline instantaneous risk of an adenoma at age $a=20$ years is given by $\psi_i(20)= \exp(\alpha_{0i} + \alpha_1 \mbox{female}_i$) where  $\alpha_{0i} \sim N(A,\sigma_\alpha)$ and $\alpha_1$ captures the difference in risk for women (female$_i=1$ indicates agent $i$ is female). Adenoma risk changes over time, generally increasing with age, a process we model using a linear change-point for log-risk with knots at ages 50, 60, and 70. 
\begin{equation}\label{adenomariskeqn}
\begin{aligned}
log(\psi_i(a)) =  
\alpha_{0i} + \alpha_1 \mbox{sex}_i   
&+ \delta(a\geq 20)\min(a-20,30)\alpha_{20}  \\
&+ \delta(a\geq 50)\min((a-50),10)\alpha_{50}  \\
&+ \delta(a\geq 60)\min((a-60),10)\alpha_{60}  \\
&+ \delta(a\geq 70)(a-70)\alpha_{70} 
\end{aligned}
\end{equation}

\subsection{Adenoma Growth Model} 
For each adenoma, we simulate a hypothetical time to reach 10mm, $t_{10mm}$, which may exceed the  agent's lifespan. We assume that $t_{10mm}$ has a Fr\`echet 
%(or type II extreme value) 
distribution with shape parameter $\beta_1$, scale parameter $\beta_2$, and cumulative distribution function given by
\begin{equation}\label{timeto10eqn}
F(t) = \exp \left[ - \left( \frac{t}{\beta_2} \right)^{-\beta_1} \right]
\end{equation}
for $t \ge 0$, with $E(t_{10mm})= \beta_2\Gamma(1-1/\beta_1)$ and median$(t_{10mm}) = \beta_2\ln(2)^{-1/\beta_1}$.
%The Fr\`echet distribution has a long right tail, with skewness that can persist as the mean moves away from zero, allowing a large proportion of slow growing adenomas. 
 Prior distributions for adenoma growth parameters specify that most adenomas grow very slowly. We allow different scale and shape parameters for adenomas in the colon and rectum.

Adenoma size at any point in time is simulated using a von Bertalanffy growth curve model
 \citep[][see also \S \ref{sec:modelappendix}]{tjorve2010unified}. The simulated time to reach 10mm is used in combination with the growth curve model to calculate the adenoma growth rate parameter. 

\subsection{Model for Transition from Adenoma to Preclinical Invasive Cancer}
For the $j$th adenoma in the $i$th agent the size at transition to preclinical cancer (in mm)  is simulated using a lognormal distribution; the underlying (exponentiated) normal distribution is assumed to have standard deviation 0.5 and mean 
\begin{align}
\begin{split}
\mu_{ij} = 
\quad &\gamma_0 + 
\gamma_1 \mbox{female}_{i} + 
\gamma_2 \mbox{rectum}_{ij} + 
\gamma_3 \mbox{female}_{i} \mbox{rectum}_{ij} + \\
\Bigl( 
&\gamma_4 + 
\gamma_5 \hbox{female}_{i} + 
\gamma_6 \mbox{rectum}_{ij}  + \gamma_7 \mbox{female}_{i} \mbox{rectum}_{ij}
\Bigr) 
\mbox{age}_{ij}.
\end{split}
\label{meantranstoca}
\end{align}
Where rectum$_{ij}$ is an indicator of rectal versus colon location and \mbox{age}$_{ij}$ is the age at adenoma initiation. %, centered at age 50.  
Based on this model, the probability that an adenoma transitions to preclinical cancer increase with increasing size. The expected size at transition is given by $\exp(\mu_\gamma + 0.125)$, with median $\exp(\mu_\gamma)$ and variance $0.28\exp(2\mu_\gamma + 0.25)$. %Negative parameter values indicate that transition tends to occur at smaller adenoma sizes. Under this model, 
Most adenomas do not reach transition size and small adenomas are unlikely to transition to cancer. For example, if $\mu_\gamma=3.5$ then the probability of transition to preclinical cancer is less than $1 \times 10^{-5}$ at 10mm, 0.008 at 15mm and 0.16 at 20mm. %Prior calibration provided little information about differential effects of rectal location on the probabiIity of transition, and so we specify uniform prior distributions for  parameters $\gamma_2$ and $\gamma_6$.

\subsection{Model for Sojourn Time}
Sojourn time is the time from the transition to preclinical (asymptomatic) CRC and clinical (symptomatic and detected) cancer. %The CRC-SPIN~2.0 model uses an updated sojourn time model to allow a longer sojourn time so that the model can more accurately reproduce cancers detected at screening. 
We simulate sojourn time using a Weibull distribution with shape parameter fixed at 5: 
\begin{equation}\label{mst}
f(x) = \Big(\frac{5}{\tau}\Big)\Big(\frac{x}{\tau}\Big)^{4}\exp\Big(-\Big(\frac{x}{\tau}\Big)^5\Big)
\end{equation}
so that $E(x)=\tau\Gamma(1.2)$ and $Var(x)=\tau^2\Big(\Gamma(1.4)-\Gamma(1.2)^2\Big)$. By fixing the shape parameter, we focus on distributions with a limited range of skewness to disallow distributions with heavy right tails while retaining enough flexibility to model plausible sojourn time distributions. We allow different values of $\tau$ for cancers in the colon and rectum. Prior distributions for sojourn time parameters allow the mean (and standard deviation) of sojourn time to range from 1.4 (sd 0.32) to 6.4 years (sd 1.5). 

\subsection{Simulation of Lifespan and Colorectal Cancer Survival}
Once a cancer becomes clinically detectable, we simulate stage and size at clinical detection and survival. Stage and tumor size at clinical detection are based on SEER data from 1975 to 1979, prior to diffusion of CRC screening \citep{SEER}. 
Simulated survival time after CRC diagnosis is based on a Cox proportional hazards model, estimated using SEER data from individuals diagnosed with CRC from 1975 through 2003 \citep{rutter2013secular}. CRC survival is based on the first diagnosed CRC and depends on sex, age at diagnosis, cancer location (colon or rectum) and stage at diagnosis.  

Other-cause mortality is modeled using survival probabilities based on product-limit estimates for age and birth-year cohorts from the National Center for Health Statistics Databases \citep{NCHS}. 

\section{Calibration Data}\label{sec:calibrationdata}
Calibration data are derived from published studies, and typically take the form of summary statistics with known distributions, such as binomial, multinomial, and Poisson. 
%We use two types of data to calibrate model parameters: individual-level data and adenoma-level data. In total, w
We calibrate to 37 targets from six sources: SEER registry data \citep[16 targets, \S \ref{sec:SEER}]{SEER} and five published studies (21 targets, \S \ref{sec:epistudies}). We also 
%incorporated information about adenoma growth from a recent study that examined individuals with two screening colonoscopies that were approximately ten years apart \citep{ponugoti2017yield}. The second screening colonoscopy detected advanced adenomas  (defined as adenomas $\geq 10$mm in size, or with villous features or high-grade dysplasia) in 3\% of individuals overall and in 9\% of individuals with at least one adenoma. Based on this, we 
bounded adenoma growth parameters, based on information from a recent study of repeated screening colonoscopies \citep{ponugoti2017yield}, 
so that the probability of an adenoma reaching 10mm within 10 years ranged from 0.0001 to 0.25, by requiring $10(-\ln(0.25))^{1/\beta_{1}} \leq \beta_{2} \leq 10(-\ln(0.0001))^{1/\beta_{1}}$.

Calibration targets are based on individual-level data that is reported in aggregate.
%, including information from published studies and registries. 
Calibration requires simulating targets by simulating a set of agents with risk that is similar to the study population based on age, gender, and prior screening patterns, and the time period of the study, which may affect both overall and cancer-specific mortality. %In general, studies describe their samples by reporting the percentage of men and women, their average age, and the standard deviation of age. Unless information was provided by sex, we assumed the same age distribution for men and women. We simulated age at the time of the study using a truncated normal distribution to match study-reported means, standard deviations, and age ranges. %We used a grid search to select a truncated normal distribution with a mean and standard deviation closest to the observed values, based on a simple distance measure.   

\subsection{SEER Registry Data}\label{sec:SEER}
SEER colon and rectal cancer incidence rates in 1975-1979 are a key calibration target (Table~\ref{SEER}). Incidence rates reported are per 100,000 individuals. These rates are based on the first observed invasive colon or rectal cancer during the years 1975-1979, the most recent period prior to dissemination of CRC screening tests. We assume that given the SEER population size, the number of incident CRC cases in any year follows a binomial distribution.  

To simulate SEER incidence rates, we generate a population of individuals from 20 to 100, with an age- and sex-distribution that matches the SEER 1978 population 
(to capture risk-levels within each age category), 
who are free from clinically detected CRC. Model-predicted CRC incidence is based on the number of people who develop CRC in the next year.

\begin{table}
	\centering
	\caption{Observed and Predicted Annual Incidence of Clinically Detected Cancers in 1975-1979, per 100,000 individuals. }\label{SEER} 
	 \begin{tabular}{|lc|rrc|rc|}
		\hline		
		 		&&&Observed&Tolerance&\mc{2}{c|}{Posterior Predicted} \\
		Location&Gender&Age&\mc{1}{c}{Mean}&Interval&\mc{1}{c}{Mean}&95\% CI \\ \hline
		Colon &Female & 20-49  &   4.8 &  (2.8, 6.8)   &   3.5 &   (2.8, 4.8) \\
			  &       & 50-59  &  43.3 & (31.3, 55.2)  &  46.3 &  (37.0, 54.2) \\
			  &       & 60-69  & 100.7 & (79.7, 121.7) & 106.0 & \phantom{1}(89.8, 119.9) \\
			  &       & 70-84  & 216.7 &(185.6, 247.8) & 210.1 & (187.7, 239.3)  \\ \hline		
		Colon &Male   & 20-49  &   4.5 &  (2.5, 6.5)   &   3.4 &   (2.6, 4.7)  \\  
			  &       & 50-59  &  45.9 & (33.2, 58.6)  &  51.0 &  (41.4, 58.1) \\
			  &       & 60-69  & 121.4 & \phantom{1}(96.6, 146.2) & 126.2 & (107.0, 143.5) \\
			  &       & 70-84  & 268.4 &(224.6, 312.2) & 261.6 & (228.7, 301.4) \\ \hline
		Rectal&Female & 20-49  &   1.9 &  (0.6, 3.1)   &   1.7 &   (0.7, 2.8)  \\
			  & 	  & 50-59  &  20.4 & (12.2, 28.6)  &  20.4 &  (13.8, 27.3) \\
			  &       & 60-69  &  42.5 & (28.9, 56.1)  &  41.9 &  (31.7, 53.1) \\
			  &       & 70-84  &  73.9 & (55.7, 92.1)  &  73.2 &  (58.1, 89.7)    \\ \hline
		Rectal&Male   & 20-49  &   2.3 &  (0.9, 3.7)   &   2.4 &   (1.3, 3.5)  \\  
			  &       & 50-59  &  30.0 & (19.7, 40.3)  &  31.7 &  (23.1, 39.5) \\
			  &       & 60-69  &  71.4 & (52.4, 90.4)  &  67.9 &  (54.5, 83.4)  \\
		      &       & 70-84  & 128.0 & \phantom{1}(97.7, 158.3) & 120.5 & (100.0, 146.9) \\ \hline

	\end{tabular}
\end{table}  

\subsection{Other Published Targets}\label{sec:epistudies}
Table~\ref{calibrationstudies} summarizes calibration targets from five studies. To simulate these targets, we generated separate populations for each target that match the age and gender distribution of study participants during the time-period of the study. One study \citep{Church} describing the pathology of lesions (i.e., adenomas and preclinical cancers) did not provide information about the age or sex of patients, and so we simulated a population that was 50\% male with an average age of 65 (standard deviation of 5), and an age range of 20 to 90 years. 

Simulation of targets in Table~\ref{calibrationstudies} also requires simulating the detection of lesions (adenomas and preclinical cancers).  Sensitivity is a function of lesions size, and is informed by back-to-back colonoscopy studies \citep[additional details provided in \S A]{Hixson,Rex97}.
We assume that study participants are free from symptomatic (clinically detectable) CRC and have not been screened for CRC prior to the study. This is a reasonable assumption because studies used for model calibration were conducted prior to widespread screening, or were based on minimally screened samples.
CRC screening guidelines have been in place since the late 1990s \citep{winawer1997colorectal}, and screening rates have since risen steadily \citep{meissner2006patterns,centers2011vital}. 
   
  \begin{table}
  	\begin{centering}
  	\caption{Observed and Predicted Calibration Targets from Published Studies }\label{calibrationstudies} 
  	\begin{tabular}{|lrc|rc|}
  		\hline
  		&&Tolerance&\mc{2}{c|}{Posterior Predicted} \\
  		Target &\mc{1}{c}{Mean}&Interval&\mc{1}{c}{Mean}&95\% CI \\ \hline  		
  		\cite{Corley2013}&&&& \\
  		\hspace*{.1in}Adenoma Prevalence, Women 50-54 & 15\phantom{.0} &(12.9, 20.8) &16.8 &(14.1, 19.9)\\
  		\hspace*{.1in}Adenoma Prevalence, Women 55-59 & 18\phantom{.0} &(15.5, 25.0) &20.3 &(17.4, 23.6)\\
  		\hspace*{.1in}Adenoma Prevalence, Women 60-64 & 22\phantom{.0} &(19.4, 30.1) &23.8 &(20.6, 27.1)\\
  		\hspace*{.1in}Adenoma Prevalence, Women 65-69 & 24\phantom{.0} &(20.6, 33.4) &27.0 &(23.5, 30.4)\\ 
  		\hspace*{.1in}Adenoma Prevalence, Women 70-74 & 26\phantom{.0} &(21.5, 37.0) &29.9 &(26.1, 33.4)\\
  		\hspace*{.1in}Adenoma Prevalence, Women $\geq$75 & 26\phantom{.0} &(20.8, 37.7) &33.2 &(29.0, 37.2)\\ \hline
  		
  		\hspace*{.1in}Adenoma Prevalence, Men 50-54      & 25\phantom{.0} &(22.1, 34.2) &26.0 &(22.7,29.6)\\
  		\hspace*{.1in}Adenoma Prevalence, Men 55-59      & 29\phantom{.0} &(25.6, 39.7) &30.7 &(26.8,34.6)\\
  		\hspace*{.1in}Adenoma Prevalence, Men 60-64      & 31\phantom{.0} &(27.5, 42.3) &35.1 &(30.9,39.3)\\
  		\hspace*{.1in}Adenoma Prevalence, Men 65-69      & 34\phantom{.0} &(29.6, 46.9) &39.2 &(34.4,43.8)\\ 
  		\hspace*{.1in}Adenoma Prevalence, Men 70-74      & 39\phantom{.0} &(33.2, 54.6) &42.7 &(37.4,47.7)\\
  		\hspace*{.1in}Adenoma Prevalence, Men $\geq$75   & 38\phantom{.0} &(31.6, 53.9) &46.6 &(40.4,52.1)\\ \hline\hline 
  		
   		\citet{Pickhardt}$^*$&&&& \\
%  		\hspace*{.1in}Percent of Detected Adenomas $\leq 5$mm         &62.0 &(55.3, 68.8) &68.5 &(66.2, 70.4)\\
%  		\hspace*{.1in}Percent of Detected Adenomas $6-9$mm     &28.7 &(22.4, 35.0) &17.1 &(15.8, 18.6)\\
  		\hspace*{.1in}Percent of Detected Adenomas $\ge 10$mm     & 9.2 &\phantom{1}(5.2, 13.2) &12.2 &(10.7, 13.2)\\ \hline\hline 
  		\citet{Imperiale00}&&&& \\
  		\hspace*{.1in}Detected Preclinical Cancers per 1,000 People &6.0 &\phantom{11}(0.3, 117.1) &2.4 &(1.8, 5.3) \\ \hline \hline  
  		\citet{Lieberman2008}$^*$&& && \\
  		\hspace*{.1in}Preclinical CRCs per 1,000 Lesions $6-9$mm     & 2.5 &(0.0, 8.4)   &4.7 &(2.1, 7.6)\\
  		\hspace*{.1in}Preclinical CRCs per 1,000 Lesions $\geq 10$mm &32.8 &(11.6, 54.0) &41.4&(29.2, 52.9)\\                      
  		\hline\hline
  		\citet{Church}&&&& \\
  		\hspace*{.1in}Preclinical CRCs per 1,000 Lesions $[6,10)$mm  & 2.4 &\phantom{1}(0.0, 10.3) &5.6 &(2.5,9.0)\\
  		\hspace*{.1in}Preclinical CRCs per 1,000 Lesions $\geq 10$mm &42.3 &(12.6, 72.1) &36.7 &(24.3, 49.0)\\
  		\hline 			
  	\end{tabular}
  \end{centering}
    $^*$Size was reported categorically as $\leq 5$mm, 6 to 9mm, and $\geq10$mm. We operationalized these categories as: $[1,5.5)$ mm, $[5.5,9.5)$ mm and $\geq 9.5$ mm
  \end{table}

%   Data from a case series \citep{Church}, describes the pathology of lesions (i.e., adenomas and preclinical cancers) detected by colonoscopy and removed between January 1995 and September 2002 in a single endoscopist's practice.  
   %We calibrated to preclinical cancer rates in lesions  $\ge 6$mm because very few cancers were detected in lesions $< 6$mm (2 in 2066).  Among 418 lesions 6 to 10mm, 1 cancer was detected (.24\%) and among 496 lesions 10mm or larger 21 adenomas were detected (4.2\%). 
%   Because this study did not provide information about the age or sex of patients, we simulated a population that was 50\% male with an average age of 65 (standard deviation of 5), and an age range of 20 to 90 years. 

\section{Posterior Inference via Incremental Mixture Approximate Bayesian Computation (IMABC)}\label{sec:iabc}
The basic rejection-based ABC algorithm \citep{Tavare,Pritchard} generates model parameter vectors $\theta$ from the prior distribution, $\pi(\theta)$, then uses the model to simulate data, ${y^*}$. Draws that result in simulated data that are similar to observed data, ${y}$, are accepted. Similarity between ${y^*}$ and ${y}$ is based on user-defined summary statistics, %$f(y)$, 
a distance metric, %, $\rho$, 
and a tolerance level %$\epsilon$.  
that defines the distance of acceptable points.

In practice, simulating $\theta$ from the prior distribution can be very inefficient because the prior and posterior distributions are often poorly aligned. 
%and this leads to proposed values that are located in low posterior probability regions. 
Many versions of ABC have been developed to address inefficiencies. Two popular variants are ABC-MCMC \citep{Marjoram} and sequential Monte Carlo ABC  \citep[ABC-SMC,][]{Sisson,Toni}. 
ABC-MCMC involves proposing a new value of $\theta$ by sampling $u$ from a user-specified jumping distribution, $q(\cdot)$, that is centered at zero with $\theta^{(t+1)}=\theta^{(t)}+u$. 
If simulated data based on $\theta^{(t+1)}$ are within tolerance levels for observed data  %($\rho(f({y}),f({y^*}))>\epsilon$),
then, similarly to MCMC, $\theta^{(t+1)}$ is accepted with a probability equal to the minimum of 1 and  $\frac{q(\theta^{(t+1)}\mid\theta^{(t)})\pi(\theta^{(t+1)})}
	  {q(\theta^{(t)}\mid\theta^{(t+1)})\pi(\theta^{(t)})}$. 
Drawbacks of ABC-MCMC include the usual problems with MCMC, such as correlated samples, low acceptance rates, the possibility of getting stuck in low posterior probability regions, and slow mixing requiring simulation of very long chains. 
ABC-SMC is based on importance sampling with the prior used as the proposal distribution. ABC-SMC starts by simulating a set of draws from the prior distribution.
Each subsequent set of draws %,  $\theta_{1}^{(t)},\dots,\theta_{N}^{(t)}$, 
is simulated by drawing an (importance) weighted sample from the previous set of draws and for each sampled point adding a random deviate $u$ that is drawn from a user-specified jumping distribution.  For each sampled point this process is repeated until the perturbed point is accepted (i.e., falls within the tolerance interval).
When using the ABC-SMC approach, users specify the total number of iterations, $T$, and a sequence of $T$ increasingly stringent tolerance intervals, which require accepted points to be nearer to targets as the algorithm proceeds.
After $T$ iterations, draws from the posterior distribution are simulated by drawing a weighted sample of $\theta$'s using final importance weights that are based on the sequence of jumping distributions.  %As the tolerance level is reduced, the approximate posterior distribution should converge towards the true posterior distribution. 
The population Monte Carlo ABC algorithm (ABC-PMC) is closely related to ABC-SMC, and also draws on  importance sampling \citep{Beaumont,Marin2012}. ABC-PMC uses a multivariate normal jumping distribution with covariance matrix that is based on prior draws.
%and estimates a random walk scale (that is also the kernel bandwidth) based on those simulations. 
%new points sampled via a component-wise independent random walk (a normal distribution centered at a previously accepted point chosen using importance weights). 
%This effectively yields a normal mixture sampling distribution. 

In general, ABC and its variants can be impractical or can fail when the parameter space is high dimensional, or there are many summary statistics that the simulated data must approximate  \citep{Blum}. %As discovered by \citet{McKinley}, this can preclude their application to MSM calibration. 
We propose a new ABC approach that we call incremental mixture approximate Bayesian computation (IMABC), which is well-suited to MSM calibration which involves both high dimensional parameter spaces and many calibration targets.
%The validity of ABC algorithms, in the sense that they result in samples from the approximate posterior distribution $f(\theta\mid \rho(y,y^*)<\epsilon)$, relies on the validity of the corresponding exact algorithms (e.g., see arguments made by \citet{Sisson}). 
%For instance, \citet{Sisson} write that the validity of their ABC algorithm is ``derived by construction from the validity of the combination of general SMC methods and the PRC process.'' 
%IMABC is an approximate Bayesian version of adaptive importance sampling \citep[SIR;][]{Rubin:SIR}, similar to %IMIS\citep{Steele_jcgs2006,Raftery:Bao}, with samples  
IMABC is an approximate Bayesian version of adaptive importance sampling, similar to IMIS \citep{Steele_jcgs2006,Raftery:Bao}, with samples 
drawn from the parameter space using a proposal distribution that is a mixture of normal distributions. 
Posterior estimates are based on accepted draws that are weighted to account for differences between the prior and proposal distrbutions.  
IMABC is most similar to the ABC-PMC approach \citep{Beaumont}. 
%ABC-PMC draws on ideas of importance sampling. At each iteration the proposal distribution is updated, using a normal distribution that is centered at a previously accepted point that is chosen with probability given by ratio of the prior distribution and the proposal distribution at the prior iteration. At each iteration, ABC-PMC uses a kernel approximation to the target distribution based on previous iterations and estimates a random walk scale(that is also the kernel bandwidth) based on those simulations \citep{Marin2012}. New points are sampled via a component-wise independent random walk, a normal distribution centered at a previously accepted point chosen using importance weights. a normal distribution centered at one of the previously sampled points with variance equal to twice the weighted empirical variance of the most recent set ot samples. This effectively yields a normal mixture sampling distribution. The IMABC approach we propose differs from ABC-PMC in how new points are sampled. In particular,
IMABC adds new points in regions near a subset of points that produce simulated targets closest to observed targets, whereas ABC-PMC samples points based on an approximation to the joint distribution using importance weights. 

\subsection{The IMABC algorithm}
The IMABC algorithm begins with a rejection-sampling ABC step, and updates this initial sample by adding points near a set of ``best'' points that result in simulated targets that are closest to corresponding observed targets.

Let $O_1,\dots,O_J$ denote the $J$ calibration targets, which we assume are summary statistics.  We specify tolerance bounds around targets based on $(1-\alpha_j)\times 100$\% confidence intervals, for $j=1,...,J$. Let $\alpha = (\alpha_1,\alpha_2,\ldots,\alpha_J)$.  The IMIS algorithm updates tolerance intervals so they become more stringent in later iterations. Let $\alpha^{(0)}$ be the alpha-levels used for tolerance intervals for the initial ABC step, $\alpha^{(t)}$ are alpha-levels for the $t$th iteration, and $\alpha^{*}$ are the final (user-specified) alpha-levels, corresponding to convergence of the IMABC algorithm. When searching a high dimensional parameter space, it is practical to begin with very wide tolerance intervals, corresponding to small values of $\alpha$. Final alpha-levels used to calculate tolerance intervals may vary across targets depending on the quality of and confidence in calibration targets. 

Let $S_{ij}$ denote the $j$th simulated target (corresponding to $O_j$) for the $i$th sampled point, and let $\delta_j(\theta_i,\alpha_j)=1$ if $S_{ij}$ falls within the $(1-\alpha_j)\times 100$\% confidence interval for target $O_j$. We use an intersection criterion for acceptance \citep{Conlan,Ratmann}, with $\theta_i$ is accepted when all $S_{ij}$ lie within ABC tolerance bounds, so that $\delta(\theta_i,\alpha)=\prod_{j=1}^J \delta_j(\theta_i,\alpha_j)$.

At the first IMABC step, a sample of $N_0$ points
%, $\theta_{1}, \hdots , \theta_{N_0}$, 
is drawn from the prior distribution of model parameters, $\pi(\theta)$. The algorithm then enters an updating phase. The $(t+1)$st iteration in the IMABC algorithm proceeds as outlined below: 

\vspace*{.1in}
\noindent {\em Step 1: Identify the best points and sample new points nearby}
\begin{description}
	\item[\normalfont{1A.}] Calculate p-values, $\rho_{ij}$, for each accepted $\theta_i$, based on two-sided tests of H$_0$: $S_{ij} = O_j$ versus H$_A$: $S_{ij} \neq O_j$ for $j=1,  \hdots , J$, treating $S_{ij}$ as fixed and $O_j$ as estimated with error. Often, as in our application, $O_j$ is a summary statistic, and is approximately normally distributed. We summarize model fit across multiple targets with  $\rho_{i\cdot} = \min_i(\rho_{ij})$, the worst fit across the $J$ targets. 
%It is possible to use other summaries when reasonable for the application, such as $\rho_{i\cdot} = \hbox{mean}_i(\rho_{ij})$).

%It is possible, especially for initial draws from the prior, for $\rho_{i\cdot}$ to be equal to the minimum acceptable $\alpha$-level for all accepted points. To better direct the algorithm, 
	\item[\normalfont{1B.}] Select the $N^{(c)}$ points with the largest $\rho_{i\cdot}$. When there are ties,  calculate the distance between the simulated and observed targets, $d_{i\cdot} = \sum_{j:\alpha_{j} < \alpha_{j}^*}{d_{ij}}$ where $d_{ij} = (S_{ij} - O_j)^2/O_j^2$ and select points with the largest $\rho_{i\cdot}$ and smallest $d_{i\cdot}$.  

	\item[\normalfont{1C.}] Simulate $B$ new draws around each of the $\theta_{(k)}^{(t+1)}$, $k=1,\dots,N^{(c)}$ best points by sampling from a normal distribution with mean $\theta_{(k)}^{(t+1)}$ and covariance $\Sigma_{(k)}^{(t+1)}$. 
	
	Let $p$ be the dimension of $\theta$ (i.e., the number of calibrated parameters). If there are fewer than $5p$ accepted points, then $\Sigma_{(k)}^{(t+1)}$ is set to a diagonal covariance matrix with standard deviation set to half the prior distribution standard deviation for each parameter.  If there are at least $5p$ and up to $25p$ accepted points, $\Sigma_{(k)}^{(t+1)}$ is calculated using all accepted points. If there are more than $25p$ accepted points, $\Sigma_{(k)}^{(t+1)}$ is calculated using the $25p$ accepted points nearest to $\theta_{(k)}^{(t+1)}$. This means that until the algorithm accepts $25p$ points, the the same covariance matrix is used for all normal mixtures. %Once the algorithm accepts $25p$ points covariance matrices diverge.

	\item[\normalfont{1D.}] Simulate calibration targets, $S_{ij}$, for each new draw, and resimulate targets at  center points, $\theta_{(k)}^{(t+1)}$. Accept or reject new draws and previously sampled center points based on $\delta(\theta_i,\alpha^{(t)})$. Resimulation of targets at center points enables the algorithm to move away from center points with $S_{ij}$ that are, by chance, similar to $O_i$.
\end{description}

\vspace*{.1in}
\noindent {\em Step 2: Update Tolerance Intervals}

If any $\alpha_{j}^{(t)}<\alpha_{j}^*$ and there are $50p$ or more accepted points, check to see if the tolerance can be updated. Identify $i^\prime$ associated with the median $\rho_{i\cdot}$, with $d_{i\cdot}$ as a tie breaker. 
For each potentially updated tolerance level, set $\alpha_{j}^{(t+1)} = \min( \rho_{{i^\prime}j},\alpha_{j}^*)$, then update the accepted $\theta$'s, so that they are based on $\delta(\theta_i,\alpha^{(t+1)})$, removing up to half of the previously accepted points that are furthest from the targets.

\vspace*{.1in}		
\noindent {\em Step 3: Evaluate Stopping Criteria}

If $\alpha^{(t+1)}=\alpha^*$, calculate sampling weights and the corresponding 
effective sample size (ESS). Sampling weights account for sampling of points from the normal mixture rather than the prior distribution,
$w_i = \pi(\theta_i)/q_t(\theta_i)$. 
The mixture sampling distribution, $q_t$, is given by 
$ q_t = \frac{N_0}{N_t}\pi + \frac{B}{N_t}\sum_{s=1}^t \sum_{k=1}^{N^{(c)}}H^{(s)}_k $
where $H^{(s)}_k$ is the $k$th normal distribution at iteration $s$, given by $N(\theta_{(k)}^{(s)},\Sigma_{(k)}^{(s)})$, and $N_t = N_0 + N^{(c)}Bt$, the total number of draws through iteration $t$.

The ESS for the $N_{(t+1)}$ draws is $ (\sum_{i=1}^{N_{(t+1)}} w_i^2)^{-1} $, where $w_i=0$ if $\delta(\theta_i,\alpha^{(t+1)})=0$ \citep{kish1965survey,Liu}. The algorithm stops when $\hbox{ESS} \geq N_{post}$, having obtained the desired number of draws from the posterior distribution. If $\alpha^{(t+1)}=\alpha^*$ and $\hbox{ESS} < N_{post}$ the algorithm continues to iterate, but without further updates to tolerance intervals.

\vspace*{.2in}
Once the IMABC algorithm is complete, independent draws from the posterior distribution are simulated by taking a weighted sample from accepted points with replacement, using the $w_i$. Alternatively, posterior means and 95\% credible intervals (CIs) can be estimated using weighted means and percentiles based on all accepted draws.

When implementing the IMABC algorithm, we recommend using a large initial sample size, $N_0$, to ensure exploration of the parameter space and because few initially sampled points may lie in high posterior probability regions. 
%The initial tolerance level, $\alpha^{(0)}$, can be set arbitrarily small because it will be updated at subsequent draws. 
The number of normal mixtures used to draw new points at each step, $N^{(c)}$, can be selected to optimize use of computing resources.  
%$B$, the number of points drawn at each of the $N^{(c)}$ mixture distributions; and
The effective sample size of the final set of accepted points, $N_{post}$, will depend on the planned uses of the calibrated targets. For example, $2,000$ is a good choice when the goal is to provide interval estimates of model predictions based on percentile intervals, but larger samples may be desired when estimating functions of parameters. %Guidelines from the IMIS literature suggest using $N_0=1000n_{p}$ and $B=100n_{p}$ where $n_{p}$ is the number of calibrated parameters. 

Using IMABC to calibrate an MSM requires multiple model evaluations at each parameter draw and the user needs to specify $m_j$, the size of the simulated sample used to obtain $S_{ij}$. $m_j$ may be smaller for common outcomes (such as adenoma prevalence), and larger for rare outcomes (such as cancer incidence). Setting $m$ too low will result in too much stochastic variation in $S_{ij}$ and inaccurate identification of acceptable $\theta_i$. Setting $m$ too high will unnecessarily slow the algorithm. 

\section{CRC-SPIN 2.0 Calibration Results}\label{sec:results}

\subsection{IMABC Implementation}
To calibrate CRC-SPIN~2.0, we used $N_{0}=21,000$ with Latin hypercube sampling from the prior distribution to ensure coverage of the parameter space at the initial draw. With the exception of the SEER target, we began with  $\alpha^{(0)}=0.0001$ and worked toward $\alpha^*=0.001$. For SEER targets we began with  $\alpha^{(0)}=0$, accepting all points regardless of nearness to SEER targets, and worked toward $\alpha^{*}=1\times10^{-9}$, which results in narrow bands around these registry-based incidence rates. Tolerance intervals are wider for study-derived targets because of the smaller sample sizes. These wider tolerance intervals also reflect the greater uncertainty in these targets due to a range of factors related to their simulation, including uncertainty about population characteristics, sensitivity of lesion detection, and lesion size measurement and categorization. Because the \citet{Corley2013} study is based on insured patients who underwent colonoscopies from 1/1/2006 to 12/31/2008, the observed adenoma incidence rates may be lower than expected because of prior screening and removal of adenomas. Therfore, we specified asymetric tolerance limits for the \citet{Corley2013} target, extending the upper tolerance range by adding 0.25$O_j$ to the upper tolerance limit.

To take advantage of high performance computing and parallel processing (Appendix \S \ref{sec:hpc}), we used $N^{(c)}=10$, drawing $B=1,000$ points from each normal mixture so that $10,000$ new points were evaluated at each updating iteration. We assumed a normal distribution for sample statistics when estimating $(1-\alpha)\times 100$\% confidence intervals and p-values. We set the final effective sample size, $N_{post}$, to $5,000$. 

When simulating target data, we used $m_j$ equal to $5\times 10^4$ for \citet{Pickhardt}; $2\times 10^5$ for \citet{Corley2013} and \citet{Imperiale00}; $3\times 10^5$ for \citet{Church}, $5\times 10^5$ for \citet{Lieberman2008} and $5\times 10^6$ for the SEER registry data. To improve efficiency of the IMABC algorithm, we sequentially calculated $S_{ij}$ for each new $\theta_i$ in Step 1 of the algorithm, working from targets that are least to most computationally intensive. After calculating each target, we evaluated $\delta_j(\theta_i,\alpha_j)$ and once $\delta_j(\theta_i,\alpha_j)=0$, the point is rejected without simulating the remaining, more computationally intensive, targets. 

Both the IMABC algorithm and the CRC-SPIN~2.0 model were implemented in the R programming language \citep{Rlanguage}. They were coupled to produce an integrated, dynamic, high-performance computing workflow with the use of the Extreme-scale Model Exploration with Swift (EMEWS) framework~\citep{ozik2016from}. Further details about the computing environment are provided in Appendix \ref{sec:hpc}.

\subsection{Posterior Estimates}
The IMABC algorithm completed 8 iterations, obtaining $5,253$ parameter draws within tolerance limits, with an effective sample size of $5,168$ draws from the joint posterior distribution. Sampling weights ranged from
$1.3 \times 10^{-4}$ to $3.20 \times 10^{-4}$, with a mean and median of $1.9 \times 10^{-4}$. 

Posterior estimated means and 95\% CIs of model parameters were based on weighted means and percentiles of accepted draws from the joint posterior distribution (shown in Table \ref{modelsummary}). We estimated that adenoma risk is higher for men than women, increases with age, and increases more rapidly at younger (than older) ages. Parameters that govern the time for an adenoma to reach 10mm were tightly estimated, with the exception of $\beta_{1R}$. %Significant prior to posterior learning is taking place for these parameters.  
Consistent with prior limitations, the model predicted that 0.4\% of adenomas in the colon reach 10mm within 10 years (95\% CI (0.4\%, 1.0\%)) and 1.8\% of adenomas in the rectum reach 10mm within 10 years (95\% CI (0.002\%, 9.8\%)). The predicted percent of adenomas reaching 10mm within 20 years rises to 9.6\% (95\% CI (6.8\%, 12.3\%)) for adenomas in the colon and 59.7\% (95\% CI (38.8\%, 83.3\%)) for adenomas in the rectum.
We estimated that adenomas transition to preclinical cancer at smaller sizes for women, for adenomas in the rectum, and for adenomas initiated at later ages. The gender effect was stronger for adenomas in the colon than for adenomas in the rectum. We did not find evidence of differential effects of age at adenoma initiation on size at transition by adenoma location or agent sex (based on interaction terms $\gamma_5$, $\gamma_6$, and $\gamma_7$).  
We estimated shorter sojourn times for preclinical cancers in the colon relative to the rectum. The estimated posterior mean sojourn time is 1.75 years with 95\% CI (1.39, 2.44) for preclinical cancers in the colon and 2.13 with 95\% CI (1.42, 3.26) for preclinical cancers in the rectum. 

\begin{figure}	
	\centering
	\begin{subfigure}[t]{3in}
	\includegraphics[width=3in]{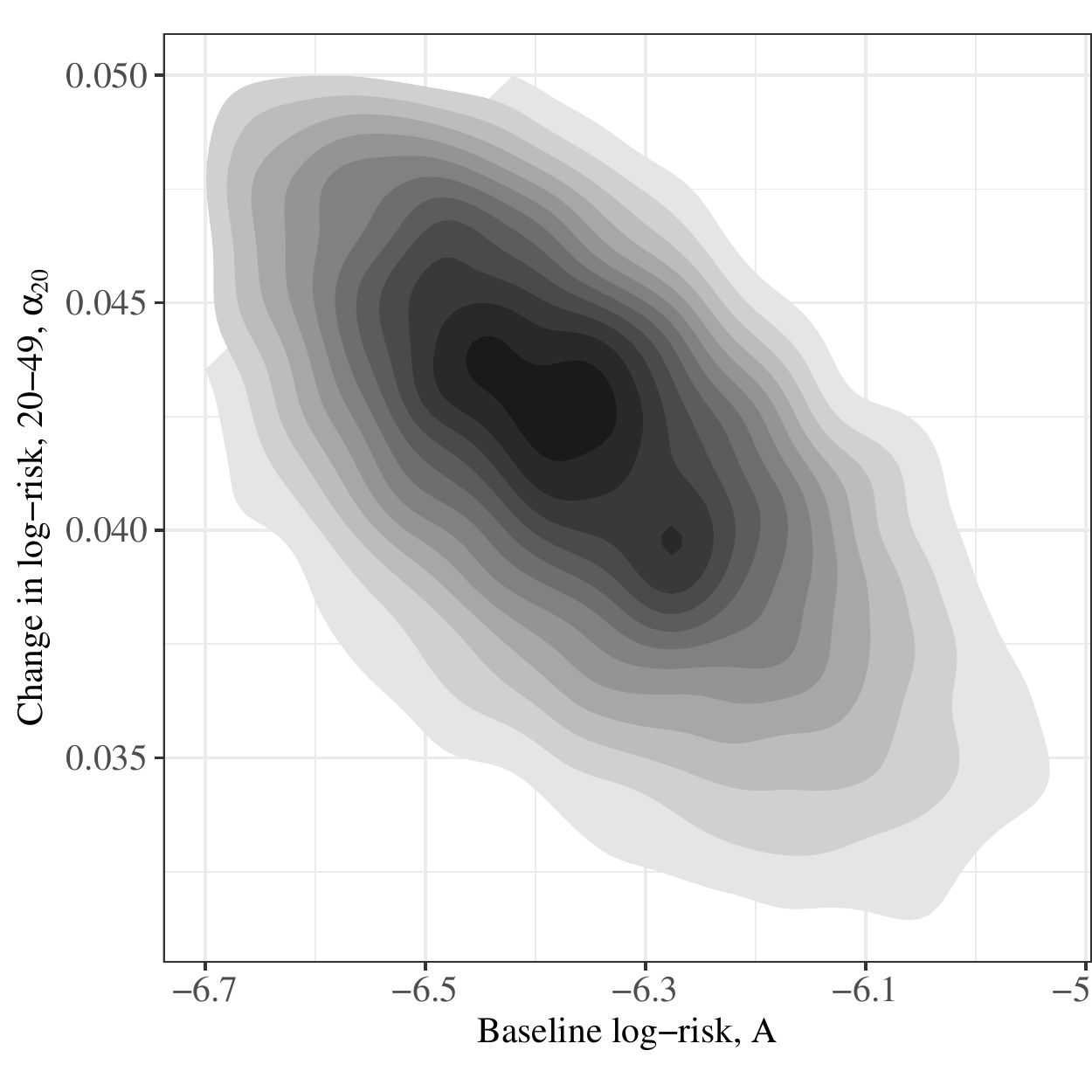}
	\caption{Baseline log-risk ($A$) and change in risk ages 20-59 ($\alpha_{20}$)}	
	\end{subfigure} 
\hspace*{0.25in}	
	\begin{subfigure}[t]{3in}
		\includegraphics[width=3in]{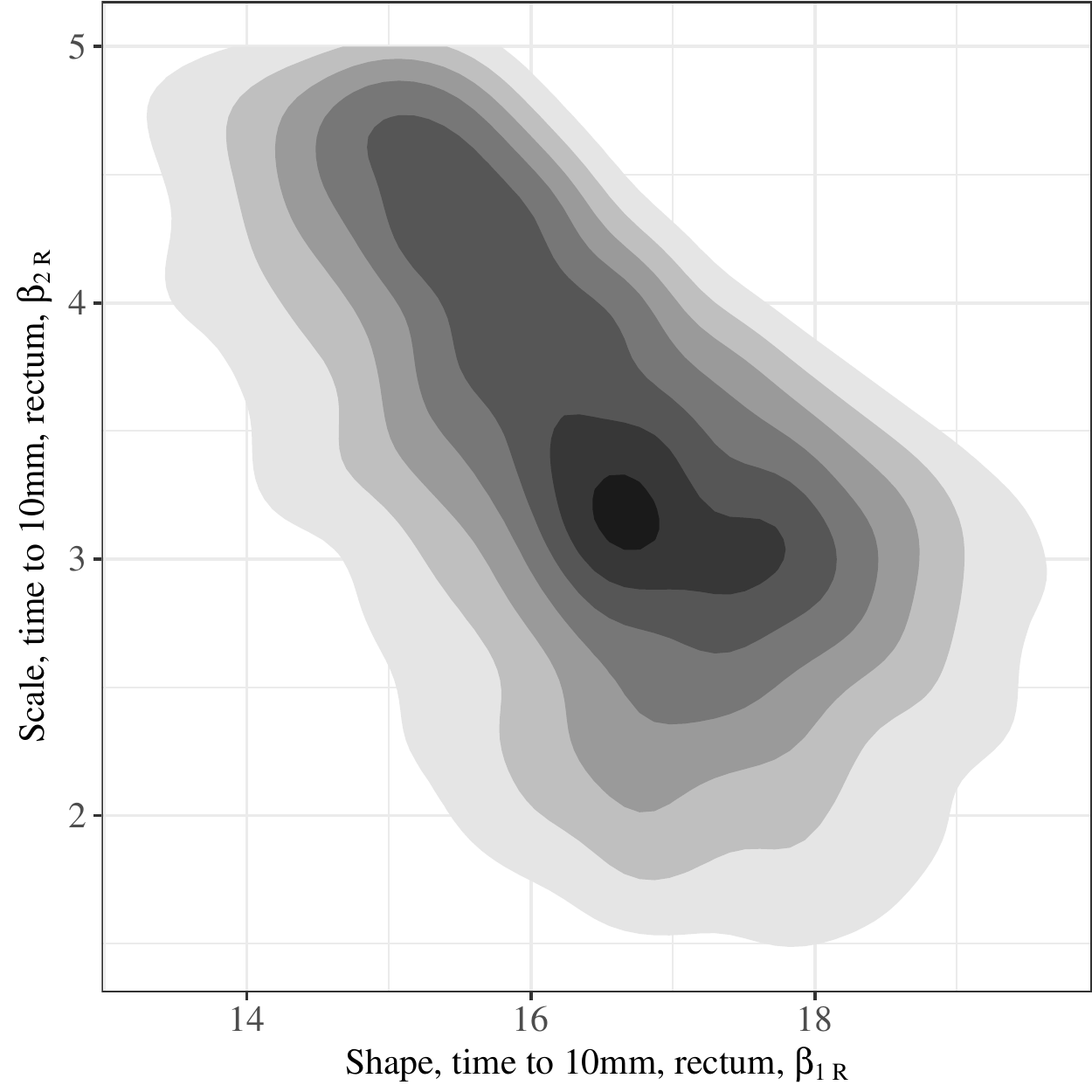}
		\caption{Growth parameters, rectal adenomas (shape, $\beta_{1R}$ and scale, $\beta_{2R}$)}	
	\end{subfigure} 
	\caption{Joint posterior distribution of model parameters associated with adenoma risk, and the growth and sojourn time in the colon.}
	\label{fig:contourplots}	
\end{figure}

By simulating draws from the posterior distribution, we were able to examine correlations and relationships among model parameters. For example, Figure \ref{fig:contourplots} displays the bivariate posterior distributions of baseline log-adenoma risk ($A$) and the annual increase in risk between the ages of 20 and 50 years ($\alpha_{20}$). When baseline risk is lower, risk increases more rapidly from 20 to 50 years to accurately predict observed adenoma prevalence, which largely is based on prevalence after age 50 when guidelines recommend initiation of CRC screening (correlation is $-0.61$). Adenoma growth parameters also show negative correlation, as demonstrated by the bivariate distribution of $\beta_{1R}$ and $\beta_{2R}$ (correlation is $-0.55$).
% While some parameters were strongly related, others had posterior distributions that exhibited little correlation (for example, $\tau_{C}$ and $\beta_{2 C}$, shown in Figure \ref{fig:contourplots}.)

The posterior predicted means of SEER targets were near observed rates and posterior 95\% CIs include SEER targets (Table \ref{SEER}).  Posterior 95\% CI do not always include other targets (Table \ref{calibrationstudies}). The model predicted higher adenoma prevalence than observed by \citet{Corley2013}, especially at older ages, acknowledging the possibility of prior screening.  The model also predicted a larger number of adenomas $\geq$10mm than observed by \citet{Pickhardt}. The probability of detecting preclinical cancer came from 3 studies, and the accuracy of model predictions demonstrates how the IMABC calibration approach combines information across potentially conflicting targets.

\section{Discussion}\label{sec:discussion}

We addressed the problem of calibrating microsimulation models by developing IMABC, an ABC algorithm 
 based on the ideas of incremental mixture importance sampling (IMIS) \citep{Steele_jcgs2006,Raftery:Bao}, an adaptive Sampling Importance Resampling algorithm \citep[SIR;][]{Rubin:SIR}.  
We illustrate our approach by calibrating CRC-SPIN~2.0, an MSM for colorectal cancer, a problem that involves a relatively high dimensional parameter space and multiple targets. 

%The overall pattern of predictions from the CRC-SPIN 2.0 model, compared to the epidemiologic study data, is one of more adenomas and preclincial cancers than observed. This is similar to the CRC-SPIN 1.0 model, with the exception of predicting more preclinical cancers. The CRC-SPIN model does not yet incorporate the sessile serrated polyp (SSP) pathway. SSPs are thought to account for at least 15\% of clinically detected CRC~\citep{ijspeert2015serrated}. The CRC-SPIN model may compensate for absense of a SSP pathway, while staying within tolerance intervals, by simulating both higher adenoma prevalence and a higher prevalence of preclinical CRC within adenomas. Clinical understanding of the SSP pathway continues to evolve, alongside a growing body of evidence to enable model development \citep{ijspeert2016prevalence} and inclusion of the SSP pathway in the next version of the CRC-SPIN model. As we extend the CRC-SPIN model, we plan to explore sequential calibration approaches that can be used to add model complexity.

%IMIS addresses the sensitivity of SIR to the proposal distribution. 
Like IMIS, the IMABC algorithm iteratively updates the proposal distribution at each iteration to obtain samples from regions of the parameter space that are consistent with calibration targets.  
The resulting mixture of normal distributions with locally adaptive covariance matrices is a very flexible distribution, and the algorithm can sample from a distribution that is multimodal to better approximate the posterior distribution.
In terms of ABC algorithms, IMABC uses a new approach to selecting tolerance levels, based on $\alpha$-levels associated with a test of equality between the simulated and observed targets, which implicitly incorporates the precision of calibration targets. IMABC also provides an automated approach to tuning these tolerance intervals, requiring users to specifiy only the initial and final values whereas ABC-SMC requires prespecification of the sequence of tolerance intervals.

Other advantages of IMABC include clear stopping rules based on the effective sample size, the ability to specify which targets are most important through final tolerance intervals, and the ability to take advantage of parallelized code. 
A limitation of the IMABC algorithm, especially as applied to MSM calibration, is that IMABC can be computationally demanding.
Evaluation of a very large number of points may be necessary, and calibration targets must be simulated for each point. The computational expense can be reduced through the ordering of target evaluations, and ceasing evaluation of a point when the first set of targets fails to fall within tolerance bounds. 
We implemented IMABC as a dynamic high-performance computing (HPC) workflow via the EMEWS framework~\citep{ozik2016from}. 
While the HPC environment was advantageous for development of the IMABC approach, we found that it was not ultimately necessary for its application.  

Future work will explore the release of publicly available code to allow others to utilize IMABC. 
In addition, because calibration to summary statistics requires simulation of a large number of  model evaluations, each with a large number of agents, we plan to explore ways to improve the efficiency of IMABC model calibration.
%, such as guidance regarding user-selected tuning parameters. 
We also plan to examine efficient approaches to parameter updating when new targets become available, and sequential calibration approaches that can be used to efficiently build from simpler to more complex models.
% For example, we plan to extend the CRC-SPIN~2.0 model to include effects of race on the development of colorectal cancer and to incorporate the sessile serrated polyp cancer pathway\citep{IJspeert2015SerratedNI}. These model extension will require both additional model parameters and new calibration targets to inform these parameters. 
%Another difficulty in applying the algorithm in practice is choosing the values of the tuning parameters. This is a focus of future work, which will involve designing simulation studies to understand the influence of different tuning parameters and aim to guide future implementations of the algorithm. Other future work includes using the calibrated MSM to estimate population-level effectiveness of interventions and predict disease outcomes. Finally, we aim to develop an efficient natural history model emulator that approximates the computationally intensive CRC-SPIN~2..

\appendix

\section{CRC-SPIN 2.0: Additional Model Information}\label{sec:modelappendix}

This appendix provides information about the CRC-SPIN 2.0 model that that may be useful for understanding the model, but is not essential to understanding the calibration approach. Complete model description can be found on the \texttt{cancer.cisnet.gov} \citep{CISNETwebsite}, in the section describing model profiles. 

\subsection{Adenoma Risk Model}
Once adenomas are initiated, they are assigned a location. The distribution of adenomas throughout the large intestine follows a multinomial distribution based on data from 9 autopsy studies \citep{Blatt1961,Chapman1963,Stemmermann1973,Eide1978,Rickert1979,Williams1982,Bombi1988,Johannsen1989,Sczepanski1992}. The probabilities associated with six sites in the large intestine (from distal to proximal) are:
P(rectum) = 0.09; P(sigmoid colon) = 0.24; P(descending colon) = 0.12; 
P(transverse colon) = 0.24; P(ascending colon) = 0.23; and P(cecum) = 0.08.
For many purposes it is important to distinguish between colon and rectal locations; more detailed location information is sometimes used for determining screening test accuracy. 

\subsection{Adenoma Growth Model} 
The diameter of the $j$th adenoma in the $i$th agent at time $t$ after onset is given by
$$d_{ij}(t) = d_\infty\left[ 1 +  \left( \left(\frac{d_0}{d_\infty}\right)^{1/p} -1 \right)\exp(-\lambda_{ij} t)  \right]$$
where $d_\infty=50$ is the maximum adenoma diameter in millimeters (mm), $d_0=1$mm is the minimum adenoma diameter, $p=3$, corresponding to the von Bertalanffy growth model, and $\lambda_{ij}$ is the growth rate for the $j$th adenoma within the $i$th agent. CRC-SPIN~1.0 specified $p=1$, corresponding to the negative exponential model, but this resulted in relatively faster early adenoma growth and too few small adenomas.  

We parameterized the growth model in terms of the time it takes for the adenoma diameter to reach 10mm to improve our ability to relate adenoma growth to observable data and clinical knowledge. The growth rate, $\lambda_{ij}$, can easily be calculated given the time to reach 10mm.

\subsection{Model for Transition from Adenoma to Preclinical Invasive Cancer}
The CRC-SPIN~2.0 model for adenoma transition is a reparameterized version of the CRC-SPIN~1.0 model for adenoma transition, restated as a regression model to better evaluate differences based on agent and adenoma characteristics. 
%For example, the main effect of rectal location in the CRC-SPIN~2.0 model is equal to $\exp(-\gamma_2)$ and the effect of age at initiation in the CRC-SPIN~2.0 model is equal to $-\gamma_4)$.
%\begin{itemize}
%\item Scale parameter: CRC-SPIN-O indicates the scale parameter as $\gamma_3$ and fixes $\gamma_3 = .5$.  CRC-SPIN-R uses the same approach, relabeling this as $\gamma_{\hbox{{scale}}}=.5$.
%\item Main effects, including the intercept: CRC-SPIN-O writes these as a scaling factors. For the CRC-SPIN-R model we re-express this, so that the main effects for CRC-SPIN-R are equal to $-\log(\mbox{main effects for CRC-SPIN-O})$.
%	\item Age effects: Age effects are reversed, so that age-effect CRC-SPIN-R is equal to -1 times age-effect CRC-SPIN-O 
%\end{itemize}

\subsection{Model for Sojourn Time}
CRC-SPIN~2.0 uses a Weibul for sojourn times. This allows longer sojourn times and better aligns with findings from previous studies than the log-normal model used in Version 1.0. For example, data from the TAMACS study \citep{TAMACS}, reported an estimated mean sojourn time of 2.85 years with a 95\% confidence interval $(2.15,4.30)$. 

\subsection{Simulation of Lifespan and Colorectal Cancer Survival}
The CRC-SPIN~2.0 model first simulates the stage at clinical detection given sex and age at detection, and then simulates size at detection conditional on stage. (In contrast, the CRC-SPIN~1.0 model simulated size, and then stage conditional on size.) 

\subsection{Simulated Screening}
Colonoscopy sensitivity for adenoma and preclinical CRC detection is based on a quadratic function of lesion size ($s$) that was successfully used in the CRC-SPIN~1.0 model. For adenomas, we assume $P(\hbox{miss}|\hbox{size}=s\leq 15\hbox{mm})=0.34 -0.0349s+ 0.0009s^2$, $P(\hbox{miss}|\hbox{size}=15 < s\leq 30\hbox{mm})=0.01$,
$P(\hbox{miss}|\hbox{size}=30 < s\leq 40\hbox{mm})=0.005$ and $P(\hbox{miss}|\hbox{size}= s\geq 40\hbox{mm})=0.001$. This function results in sensitivity that is consistent with observed findings from the 1990's \citep{Hixson,Rex97}: sensitivity is 0.76 for a 3mm adenoma, 0.87 for a 7.5mm adenoma, and 0.95 for a 12mm adenoma. For preclinical cancers, we assume sensitivity that is the maximum of 0.95 and sensitivity based on adenoma size, so that colonoscopy sensitivity is 0.95 for preclinical cancers 12mm or smaller, and sensitivity is greater than 0.95 for preclinical cancers larger than 12mm.  

Participants in the \citet{Pickhardt} study underwent both CT colonography (CTC) and colonoscopy for the purposes of evaluating the accuracy of CTC, primarily for adenomas 6mm and larger. The sensitivities reported by  \citet{Pickhardt} are consistent with those used for onetime colonoscopy.

\section{Programming and Computing Environment}\label{sec:hpc}

%In order to satisfy the computational requirements of running IMABC at the necessary scales, w
We utilized the EMEWS framework~\citep{ozik2016from} to implement a dynamic HPC workflow controlled by the IMABC algorithm. EMEWS, built on  the general-purpose parallel scripting language Swift/T~\citep{wozniak2013swiftt}, allows for the direct integration of multi-language software components, in this case IMABC and CRC-SPIN~2.0, and
can be used
% enables heuristic exploration of high-dimensional parameter spaces 
on computing resources ranging from desktops and campus clusters to supercomputers. The resulting IMABC EMEWS workflow is driven directly by the IMABC R source code, obviating the need for porting the code to alternate programming languages or platforms for the sole purpose of running large-scale computational experiments. 

The experiments were performed on the Cray XE6 Beagle at the University of Chicago, hosted at Argonne National Laboratory. Beagle has 728 nodes, each with 2 AMD Operton 6300 processors, each having 16 cores, for a total of 32 cores per node; the system thus has 23,296 cores in all. Each node has 64 GB of RAM. 
Experiments were also run on the Midway2 cluster at the University of Chicago Research Computing Center. 
Midway2 is a hybrid cluster, including both CPU and GPU resources. For this work, the CPU resources were used, consisting of 370 nodes of Intel E5-2680v4 processors, each with 28 cores and 64 GB of RAM. Swift/T, with the underlying EMEWS workflow engine, allows for the abstraction of resource specific settings (e.g., scheduler type and compute layouts) for a variety of target computing resources. Thus, once the IMABC EMEWS workflow was developed, it could be run on both the Beagle and Midway2 clusters with only minimal configuration modifications. 

The experiment reported here used 80 nodes on Beagle with 4 worker processes per node (to account for the memory footprint of CRC-SPIN~2.0) for a total of 320 worker processes, each of which could concurrently execute an individual model run.  The total compute time was 29.4 hours or 2,352 node-hours.

%%%%%% include this section if you wish to acknowledge people,
%%%%%% grant support, etc.

\section*{Acknowledgements}

This publication was made possible by financial support provided by NIH. Drs. Rutter and DeYoreo were supported by a grant from the National Cancer Institute (U01-CA-199335) as part of the Cancer Intervention and Surveillance Modeling Network (CISNET). Drs. Ozik and Collier were supported by the NIH (grants 1R01GM115839, 1S10OD018495), the NCI-DOE Joint Design of Advanced Computing Solutions for Cancer program, and through resources provided by the Computation Institute and the Biological Sciences Division of the University of Chicago, the University of Chicago Research Computing Center, and Argonne National Laboratory.  This material is based upon work supported by the U.S. Department of Energy, Office of Science, under contract number DE-AC02-06CH11357. The contents of this article are solely the responsibility of the authors and do not necessarily represent the official views of the National Cancer Institute.\vspace*{-8pt}

%%%%%% include this section only if your manuscript refers to supplementary
%%%%%% materials -- see Instructions for Authors at 
%%%%%% http://www.tibs.org/biometrics

\bibliography{calib}

\end{document}